\documentclass[twocolumn,showpacs,preprintnumbers,amsmath,amssymb,prl,superscriptaddress,longbibliography]{revtex4-1}
\usepackage{bbm}
\usepackage{mathrsfs}
\usepackage{graphicx}
\usepackage{dcolumn}
\usepackage{bm}
\usepackage{amsmath}
\usepackage{amsfonts}
\usepackage{color}
\usepackage[colorlinks=true,citecolor=blue,anchorcolor=cyan]{hyperref}
\usepackage{floatrow}

\def\be{\begin{equation}} \def\ee{\end{equation}}
\def\bea{\begin{eqnarray}} \def\eea{\end{eqnarray}}

\begin{document}

\title{Planar magnetic texture on the surface of a topological insulator}
\author{Zhaochen Liu}
\affiliation{State Key Laboratory of Surface Physics and Department of Physics, Fudan University, Shanghai 200433, China}
\author{Jing Wang}
\thanks{wjingphys@fudan.edu.cn}
\affiliation{State Key Laboratory of Surface Physics and Department of Physics, Fudan University, Shanghai 200433, China}
\affiliation{Institute for Nanoelectronic Devices and Quantum Computing, Fudan University, Shanghai 200433, China}
\author{Congjun Wu}
\thanks{wucongjun@westlake.edu.cn}
\affiliation{School of Sciences, Westlake University,  Hangzhou 310024, Zhejiang, China}
\affiliation{Institute for Theoretical Sciences, Westlake University, Hangzhou 310024, Zhejiang, China}
\affiliation{Key Laboratory for Quantum Materials of Zhejiang Province, School of Science, Westlake University, Hangzhou  310024,  China}

\begin{abstract}
We study the planar magnetic textures in an insulating magnetic film coupled to the Dirac surface state of a topological insulator. It is shown that the radial vortex with winding number $w=\pm1$ leads to the confinement of Dirac states, where an exact mapping to the Schr\"{o}dinger equation of a two-dimensional hydrogen atom is found. The fully spin polarized zero energy bound state resembles the zeroth Landau level of Dirac electrons in a uniform out-of-plane magnetic field. Interestingly, when the hybrid system is proximity coupled to an $s$-wave superconductor, the existence of Majorana zero modes at Abrikosov vortex depends only on the relative value of the magnetic exchange coupling and the pairing strength. 
We conclude with a brief discussion on the physical realization with such magnetic textures.
\end{abstract}

%\date{\today}

%\pacs{
%        73.22.-f  % Electronic structure of nanoscale materials and related systems
%        02.20.-a  % Group theory
%        73.43.-f  % Quantum Hall effects
%      }

\maketitle

\emph{Introduction.-}
Topology has become a central theme in condensed matter physics.
Interesting quantum phenomena emerge from the intricate interplay 
between non-trivial topology and magnetism \cite{hasan2010,qi2011,tokura2019,yang2020}.
Two outstanding examples are the quantum anomalous Hall
effect and axion insulators discovered in magnetic topological insulators (TI)~\cite{qi2008,yu2010,chang2013b,checkelsky2014,kou2014,bestwick2015,
deng2020,wang2015d,liu2016,wang2015b,mogi2017,mogi2017a,grauer2017,xiao2018,liu2020}. The exchange coupling between an out-of-plane magnetization and
the TI surface state opens a gap in the surface
spectrum.
The gap opening is accompanied by the emergence of the surface quantum
Hall effect with a half-quantized Hall conductance described by the axion electrodynamics~\cite{qi2008},
which is is the physical origin for the topological magnetoelectric effect.
It leads to rich phenomena such as the quantum anomalous Hall with a chiral
edge state emerging at the magnetic domain wall as well as the
topological magneto-optical effect~\cite{okada2016,wul2016,dziom2017}.

Peculiar physics emerges when the Dirac surface states couple to spatially
nonuniform magnetic textures, such as
skyrmions and domain walls ~\cite{garate2010,nomura2010,yakoyama2010,yserkovnyak2012,hurst2015,yang2020}.
In this paper, we study theoretically the two-dimensional (2D) magnetic
textures, in an insulating ferromagnetic thin film proximity-coupled to
TI surface states.
The Dirac electron with a radial magnetic vortex with
the winding number $w=\pm1$ can be exactly mapped to
the Schr\"{o}dinger equation of a 2D hydrogen atom.
The fully spin-polarized zero energy bound state resembles the zeroth
Landau level of Dirac electrons in a uniform out-of-plane magnetic field. The radial magnetic vortex acts effectively as a magnetic field along $z$ axis with $1/r$ dependence. Interestingly, when the system is coupled to an $s$-wave superconductor,
the existence of emergent Majorana bound states at the Abrikosov
vortex core depends only on the relative value of the magnetic
exchange coupling and the pairing strength.

\begin{figure}[t]
\begin{center}
\includegraphics[width=3.4in,clip=true]{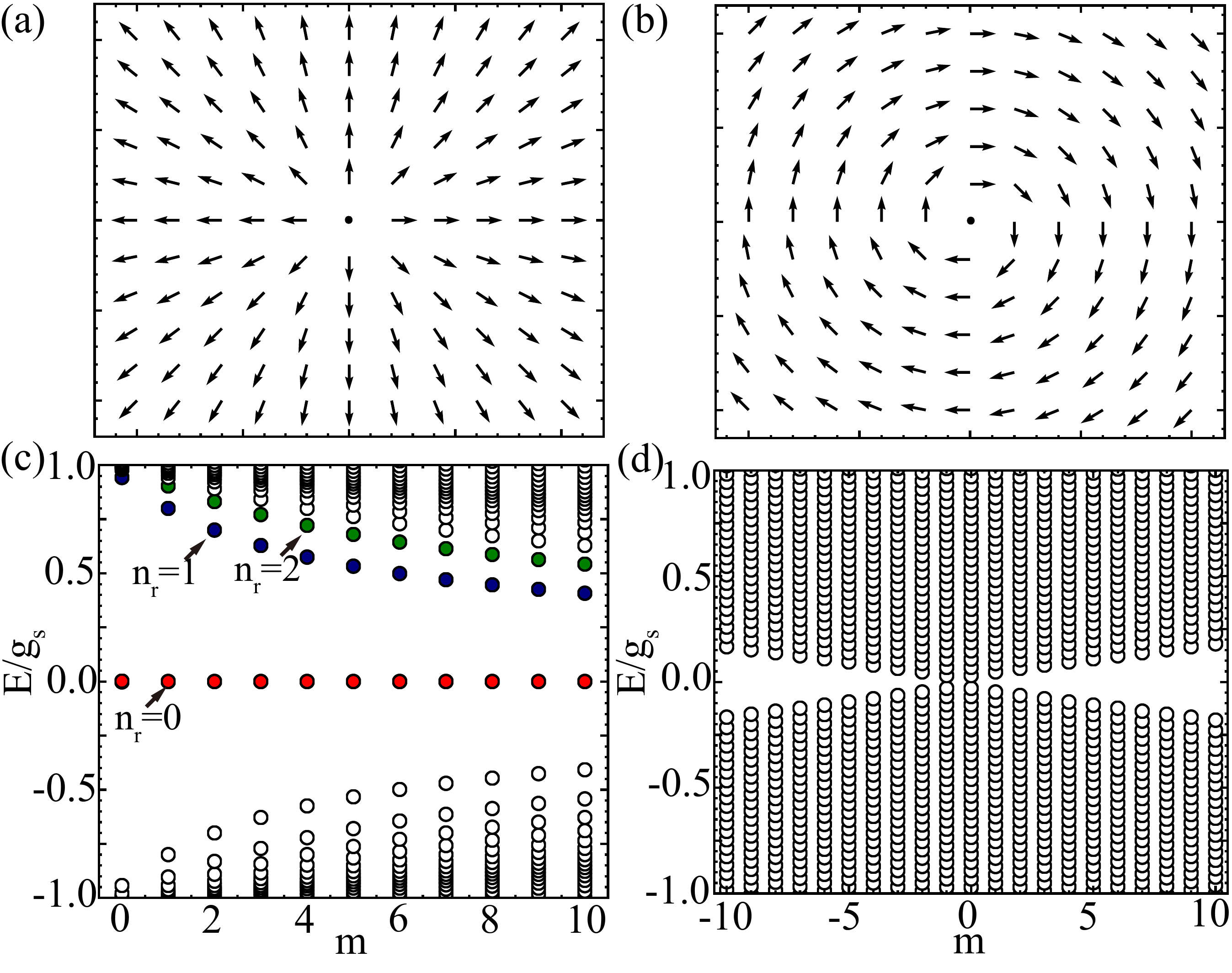}
\end{center}
\caption{(a,b) Schematics of the radial and curling planar magnetic vortices
in Eq.~(\ref{type_i}) and Eq.(\ref{type_ii}), respectively. 
The vector field represents the direction of local magnetization $\mathbf{n}(\mathbf{r})$. 
(c,d) The numerically calculated energy spectrum for the two cases in 
(a) and (b) with $g_s=2,v_F=10$, respectively. 
The bound state spectrum in (c) is bounded by $|E_\pm|<g_s$.}
\label{fig1}
\end{figure}

\emph{Model.-}
The planar magnetic texture we consider 
exhibits the form of
$\mathbf{S}(\mathbf{r})\equiv S\mathbf{n}(\mathbf{r})=S(\cos\vartheta(\mathbf{r}),\sin\vartheta(\mathbf{r}),0)$, where $\mathbf{n}(\mathbf{r})$ is the unit vector describing the magnetization direction, and $\mathbf{r}=(x,y)$. It is characterized by the topological winding number~\cite{thouless1998}
\begin{equation}
w=\frac{1}{2\pi}\oint\boldsymbol{\nabla}\vartheta\cdot d\mathbf{r}.
\end{equation}
Two typical magnetic vortices with $w=1$ are illustrated in Fig.~\ref{fig1}
(a,b).

In the presence of the magnetic vortex $\mathbf{n}(\mathbf{r})$, the surface states of a TI can be described by the Dirac Hamiltonian
\begin{equation}
\mathcal{H}=v_F(\mathbf{k}\times\boldsymbol{\sigma})
\cdot\hat{\mathbf{e}}_z+g_s\mathbf{n}(\mathbf{r})\cdot\boldsymbol{\sigma}.
\end{equation}
Here $\hbar\equiv1$, $v_F$ is the Fermi velocity, $\hat{\mathbf{e}}_z$ is the unit vector normal to the surface, and $\boldsymbol{\sigma}=(\sigma_x,\sigma_y,\sigma_z)$
are the Pauli matrices describing the spin.
$g_s\equiv J'S\rho_s/2$, $J'$ is the exchange interaction between the
local spin and surface electron, and $\rho_s$ is the sheet density
of local spin.
Without loss of generality, we set $g_s/v_F>0$ and neglect the
particle-hole asymmetry.

We notice that the in-plane Zeeman term is equivalent to a vector potential,
thus the Hamiltonian becomes
\begin{equation}\label{H}
\mathcal{H}=v_F\left[\left(\mathbf{k}-\boldsymbol{\mathcal{A}}\right)
\times\boldsymbol{\sigma}\right]\cdot\hat{\mathbf{e}}_z,
\end{equation}
where $\boldsymbol{\mathcal{A}}(\mathbf{r})\equiv (g_s/v_F)(-\sin\vartheta(\mathbf{r}),\cos\vartheta(\mathbf{r}))$.
For a generic $\boldsymbol{\mathcal{A}}(\mathbf{r})$, Eq.~(\ref{H}) is
difficult to solve analytically. In the following, we assume $\mathbf{n}(\mathbf{r})$ is rotational invariant
along $z$ axis.
The Hamiltonian conserves total angular momentum, and the energy spectrum
can be obtained analytically by squaring $\mathcal{H}$ to solve $\mathcal{H}^2\psi=E^2\psi$, which is
\begin{equation}\label{H_square}
\mathcal{H}^2/v_f^2 = -\boldsymbol{\nabla}^2+\boldsymbol{\mathcal{A}}^2+2i\boldsymbol{\mathcal{A}}\cdot\boldsymbol{\nabla}+i(\boldsymbol{\nabla}\cdot\boldsymbol{\mathcal{A}})-\boldsymbol{\nabla}\times\boldsymbol{\mathcal{A}}\cdot\boldsymbol{\sigma}.
\end{equation}
In terms of the 2D polar coordinates $(r,\phi)$, Eq.~(\ref{H_square})
is separable into the radial and angular parts.
Two typical magnetic textures are radial and curling vortices.

\emph{Radial vortex.-} First we consider a radial magnetic vortex
illustrated in Fig.~\ref{fig1}(a),
\begin{equation}\label{type_i}
\mathbf{n}_1(\mathbf{r})=\hat{\mathbf{e}}_r=\frac{1}{r}\left(x\hat{\mathbf{e}}_x+y\hat{\mathbf{e}}_y\right),
\end{equation}
where $r=\sqrt{x^2+y^2}$. Now $\boldsymbol{\mathcal{A}}_1(\mathbf{r})\equiv (g_s/v_F)(-y/r,x/r)$, and $\boldsymbol{\nabla}\cdot\boldsymbol{\mathcal{A}}_1=0$, $\boldsymbol{\nabla}\times\boldsymbol{\mathcal{A}}_1=(g_s/v_F)(\hat{\mathbf{e}}_z/r)$, $\boldsymbol{\nabla}\cdot\boldsymbol{\nabla}\times\boldsymbol{\mathcal{A}}_1=0$, and $\boldsymbol{\mathcal{A}}_1\cdot\boldsymbol{\nabla}=-(g_s/v_F)(y\partial_x-x\partial_y)/r\equiv(g_s/v_F)(iL_z/r)$, with $L_z$ the orbital angular momentum operator. Then Eq.~(\ref{H_square}) becomes
\begin{equation}
\frac{\mathcal{H}^2_{1}}{v_F^2} = -\frac{1}{r}\partial_r\left(r\partial_r\right)+\frac{L_z^2}{r^2}-\frac{g_s}{v_Fr}\left(2L_z+\sigma_z\right)+\frac{g_s^2}{v_F^2},
\end{equation}
where $\partial_r\equiv\partial/\partial r$.

Since $[L_z,\mathcal{H}_1^2]=0$, the eigenfunction must have the form as $\psi(r,\phi)=e^{im\phi}f(r)$, where the orbital momentum quantum number $m=0,\pm 1,\pm 2,...$ is connected with the total angular momentum $j=m+1/2$.
The radial wave function satisfies the equation
\begin{equation}
\label{radial_1}
\left[\frac{1}{r}\partial_r\left(r\partial_r\right)-\frac{m^2}{r^2}+\frac{g_s}{v_Fr}(2m\pm 1)-\frac{g_s^2}{v_F^2}+\frac{E^2}{v^2_F}\right]f(r)=0.
\end{equation}
Here $\pm$ denote $\left|\uparrow\right\rangle$ and $\left|\downarrow\right\rangle$ with $\sigma_z=+1$ and $\sigma_z=-1$, respectively. 
We define dimensionless quantities
\begin{equation}
r' = r\frac{g_s}{v_F}\left(m\pm1/2\right),\ \
E' = \frac{(E/g_s)^2-1}{2(m\pm1/2)^2}.
\end{equation}
Then Eq.~(\ref{radial_1}) has the exact form as the Schr\"{o}dinger equation of a 2D hydrogen atom~\cite{yang1991}, which is the main results of this paper,
\begin{equation}\label{radial_2}
\left[\frac{\partial^2}{\partial r'^2}+\frac{1}{r'}\frac{\partial}{\partial r'}-\frac{m^2}{r'^2}+\left(2E'+\frac{2}{r'}\right)\right]f(r')=0.
\end{equation}
Eq.~(\ref{radial_2}) can be solved algebraically due to the hidden dynamical symmetry of hydrogen atom~\cite{parfitt2002}, which is related to the conserved Runge–Lenz operator $\mathbf{K}\equiv-(i\boldsymbol{\nabla}\times\mathbf{L}_z-\mathbf{L}_z\times i\boldsymbol{\nabla})-2\hat{\mathbf{e}}_r$, satisfying $[\mathcal{H}^2,\mathbf{K}]=0$, $[L_z,K_x]=iK_y$, $[L_z,K_y]=-iK_x$ and $[K_x,K_y]=-4iL_z\mathcal{H}^2_1$.

Now $r'>0$ corresponds to an attractive Coulomb potential, which contains both bound discrete states and unbound continuous states. For $m\geq0$, the eigenvalue of the bound states in Eq.~(\ref{radial_1}) is $E'=-1/2n_2^2$, where the principle quantum number $n_2\equiv n_r+\left|m\right|+1/2$, with $n_r=0,1,2,...$. The radial wave function $f_{n_r,m}(r')=r'^{|m|}e^{-r'/n_2}F\left(-n_r,2|m|+1,2r'/n_2\right)$, with $F(\alpha,\gamma;x)$ the confluent hypergeometric function. The degeneracy between $\left|\uparrow\right\rangle$ and $\left|\downarrow\right\rangle$ exists when $n_{r,\uparrow}=n_{r,\downarrow}+1$ and $m_{r,\uparrow}=m_{r,\downarrow}-1$, namely $E_{+,n_r,m}=E_{-,n_r-1,m+1}$. Explicitly, the energy spectrum and (un-normalized) spinor wavefunction of bound states are
\begin{equation}
\label{bound}
\begin{aligned}
\frac{E_\pm}{g_s}&= 
\pm \left (1-\frac{(m+\frac{1}{2})^2}{n_2^2}\right)^{\frac{1}{2}},
\\
\psi_\pm &= \begin{pmatrix}
e^{im\phi}f_{n_r,m}(r'_{+})\\
\mp\frac{\sqrt{(n_r+m+1)n_r}}{(m+1)(2m+1)n_2}e^{i(m+1)\phi}f_{n_r-1,m+1}(r'_{-})
\end{pmatrix}.
\end{aligned}
\end{equation}
The bound state energies are within the interval between $\pm g_s$.
The maximum of radial probability density is located at $r_{\text{max}}\approx n_2^2a/(m\pm1/2)$, with Bohr radius $a\equiv\hbar v_F/g_s$.
For $m<0$ (and $r'<0$), Eq.~(\ref{radial_2}) corresponds to
a repulsive Coulomb potential,
there is no bound states and the continuous energy
spectrum satisfies $E'\geq0$, namely $|E|\geq g_s$.

The branch of states with $n_r=0$ are zero energy states and fully
spin-polarized as shown in their wave functions
$\psi=(e^{im\phi}f_{0,m}(r'_{+}),0)^T$.
They form a flat band resembling the spin polarized zeroth Landau level
of Dirac fermions in a uniform magnetic field.
However, the magnetic translation symmetry is absent here since
$\boldsymbol{\nabla}\times\boldsymbol{\mathcal{A}}_1$ scales as $1/r$.
Consequently, the density of states of the zero energy states scales as $1/r$
as away from the center consistent with the classic radius $r_m=(m+1/2)a$.
This situation is similar to the Landau level formation of a 2D Rashba
system subject to a harmonic potential \cite{li2012,zhou2013}.

The analytic solutions of energy spectrum is further confirmed by the numerical calculation in Fig.~\ref{fig1}(c). Here we choose a disc geometry, and the radial wave function is expanded in terms of the Bessel function~\cite{hayashi1998}, the expanding order and disc radius $\mathcal{R}$ is large enough to ensure convergence and capture the low energy states~\cite{supple}.

\emph{Curling vortex.-} Next we consider a curling magnetic vortex shown in Fig.~\ref{fig1}(b)
\begin{equation}\label{type_ii}
\mathbf{n}_2(\mathbf{r})=\hat{\mathbf{e}}_{\phi}=\frac{1}{r}(y\hat{\mathbf{e}}_x-x\hat{\mathbf{e}}_y).
\end{equation}
Then $\boldsymbol{\mathcal{A}}_2(\mathbf{r})=(g_s/v_F)(x/r,y/r)$, with a zero emergent magnetic field $\boldsymbol{\nabla}\times\boldsymbol{\mathcal{A}}_2=0$ and $\boldsymbol{\nabla}\cdot\boldsymbol{\mathcal{A}}_2=(g_s/v_F)(1/r)$. Interestingly, the curling texture can be gauged away, i.e. $\boldsymbol{\mathcal{A}}_2\rightarrow\boldsymbol{\mathcal{A}}_2'+(g/v_F)\boldsymbol{\nabla}r$ and $\psi\rightarrow\psi'\exp(igr/v_F)$. There are no bound state but only continuous state solutions~\cite{supple}. The energy spectrum is numerically calculated in Fig.~\ref{fig1}(d).

\emph{Majorana bound state.}
It is well known that the localized Majorana zero modes (MZM) arise in the Abrikosov vortex core when the Dirac fermion surface states of a TI
is proximity coupled to an $s$-wave superconductor~\cite{fu2008,hosur2011,xu2015,xu2016,wangd2018,jiang2019}.
Now we study the fate of MZM in the presense of magnetic texture on TI surface.

Now the pairing term $V=\Delta(\mathbf{r})\psi^{\dagger}_{\uparrow}\psi^{\dagger}_{\downarrow}+\text{H.c.}$ is added to $\mathcal{H}$, where $\Delta(\mathbf{r})$ is the pairing potential. The sytem can be diagonalized with the Bogoliubov transformation
\begin{equation}
\psi_{\sigma}(\mathbf{r})=\sum_{n}\left[u_{n,\sigma}(\mathbf{r})\gamma_{n}+v^{*}_{n,\sigma}(\mathbf{r})\gamma^{\dagger}_{n}\right],
\end{equation}
where $\gamma_n^\dag$ create a Bogoliubov quasiparticle. Then the resulting Bogoliubov-de Gennes (BdG) equation is
\begin{equation}\label{BdG}
\begin{pmatrix}
\mathcal{H}-\mu & \Delta(\mathbf{r})  \\
\Delta^*(\mathbf{r})  & -\sigma_y\mathcal{H}^{*}\sigma_y +\mu
\end{pmatrix}\Phi_n(\mathbf{r})=E_n\Phi_n(\mathbf{r}),
\end{equation}
with the BdG energy spectrum $E_n$ and Nambu wave function $\Phi_{n}(\mathbf{r})=\left(u_{n\uparrow}(\mathbf{r}),u_{n\downarrow}(\mathbf{r}),v_{n\downarrow}(\mathbf{r}),-v_{n\uparrow}(\mathbf{r})\right)^T$.

The BdG equation can be solved numerically for a general paring potential. Here we study the Abrikosov vortex, where the magnetic flux is contributed by the radial magnetic vortex. Thus the origin of vortex core coincides with the center of magnetic texture as in Fig.~\ref{fig3}(a), then Eq.~(\ref{BdG}) has a cylindrical symmetry. In the polar coordinate, the pairing potential with vortex is $\Delta(\mathbf{r})=\Delta_0(r)e^{i\theta}=\Delta_0\tanh(r/\epsilon_0)e^{i\theta}$,
where $\epsilon_0$ characterizes the size of vortex core. The wave function can be factorized into
\begin{equation}
\Phi_{n,\ell}=e^{i\ell\theta}\left(u_{n,\ell\uparrow},u_{n,\ell+1\downarrow}e^{i\theta},v_{n,\ell-1\downarrow}e^{-i\theta},-v_{n,\ell\uparrow}\right)^T,
\nonumber
\end{equation}
where the principle quantum number $n$ is determined by solving the radial equation in the basis of Bessel function. The details of numerical calculations are given in Supplemental Material~\cite{supple}. The calculation is performed on disc of radius $\mathcal{R}=40\epsilon_0$.

\begin{figure}[t]
\begin{center}
\includegraphics[width=3.4in,clip=true]{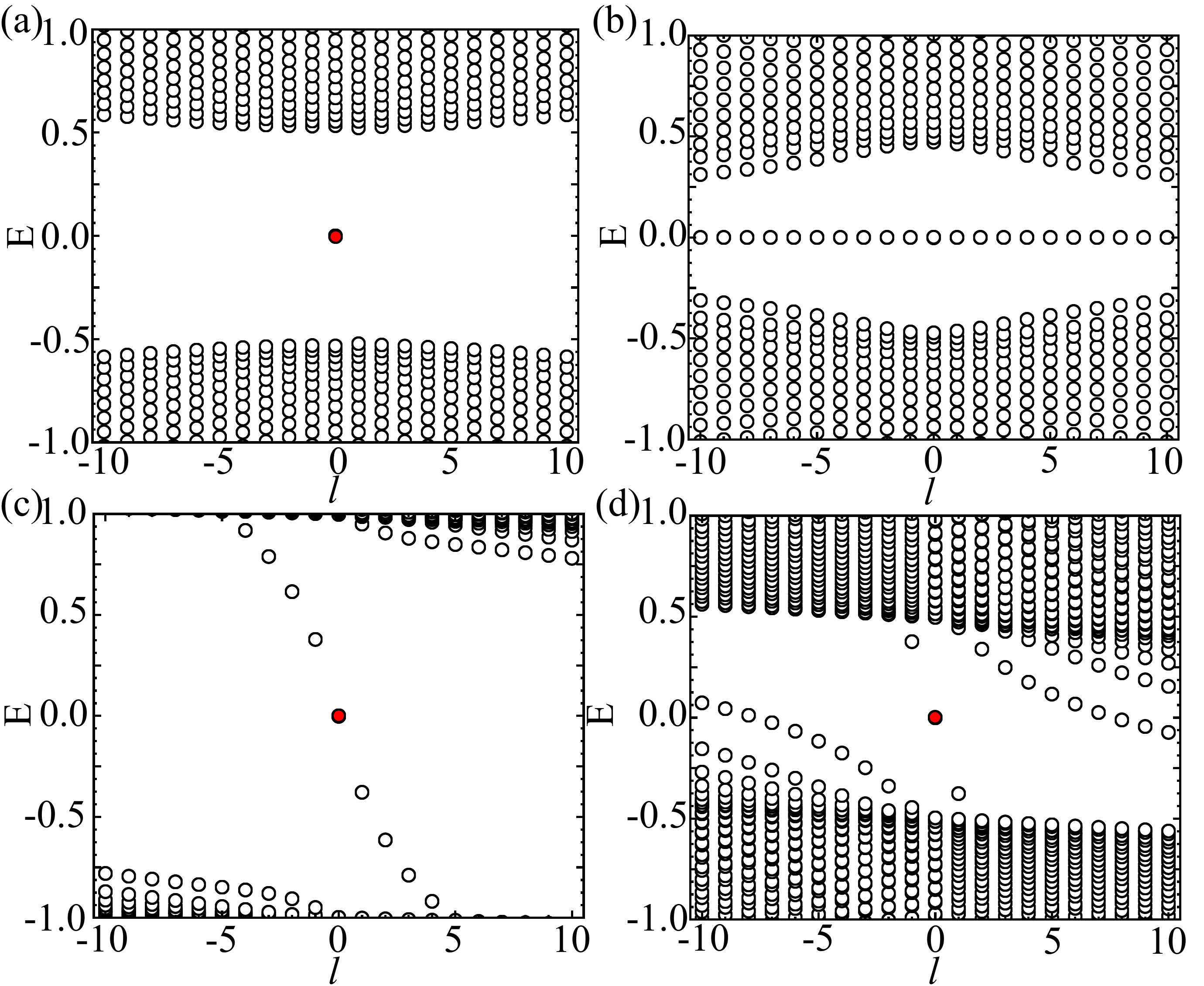}
\end{center}
\caption{The BdG energy spectrum of the Abrikosov vortex $\Delta_0(r)e^{i\theta}$ for radial vortex texture. (a) $g_s=0.5$, $\mu=0$, $\Delta_0=1.0$. (b) $g_s=1.0$, $\mu=0$, $\Delta_0=0.5$. (c) $g_s=0.5$, $\mu=5.0$, $\Delta_0=1.0$. (d) $g_s=1.0$, $\mu=5.0$, $\Delta_0=0.5$. We set $v_F=10$.}
\label{fig2}
\end{figure}

Fig.~\ref{fig2} shows the numerical results of BdG energy spectrum for a radial 
vortex. The MZM is denoted as red dot. As shown in Fig.~\ref{fig2}(a,b) for $\mu=0$, the existence of MZM depends only on the relative value of magnetic coupling $g_s$ and pairing strength $\Delta_0$, namely MZM exists for $g_s<\Delta_0$ and disappears for $g_s>\Delta_0$. The zero energy states in Fig.~\ref{fig2}(b) are not MZM, but are originated from the zero energy bound states in Fig.~\ref{fig1}(c), which deviate from zero by adding a small Zeeman term $\Delta_z\sigma_z$ into $\mathcal{H}$. 
These results can be simply understood that for $\mu=0$, 
the chemical potential resides in the bound state spectrum. 
Only when the pairing potential exceeds the bound state energy
and reaches the continuous state, then MZM from superconducting Dirac 
fermion proposed by Fu-Kane applies~\cite{fu2008}. 
This is further confirmed in Fig.~\ref{fig2}(c,d), where a finite $\mu>g_s$ 
crosses the continuous states, MZM always exists in the vortex core. 
However, for $\mu>g_s$ case, the system is gapless for finite $\ell$ 
when $g_s>\Delta_0$ shown in Fig.~\ref{fig2}(d). 
We further calculate the vortex case $\Delta_0(r)e^{i\theta}$ and $g_s<0$~\cite{supple}, where MZM exists $|g_s|<|\Delta_0|$ and disappears for $|g_s|>|\Delta_0|$. The subtle difference between positive and negative $g_s$ in the vortex case can be seen from the analytic solution. In the limit of $\epsilon_0\rightarrow0$, the vortex core (where $\Delta_0$ vanishes) can be taken to have negligible size and the boundary condition at $r\rightarrow0$ is unimportant, the analytic zero energy MZM solution is
\begin{equation}
\gamma^{\dagger}_{0,+}=(1,0,0,1)^{T}\exp\left(-\int^r_0 dr'\frac{\Delta_0(r')+g_s}{v_F}\right).\\
\end{equation}
The solution is unphysical at $r\rightarrow\infty$ when $g_s<-\Delta_0<0$. For an anti-vortex $\Delta_0(r)e^{-i\theta}$, the MZM solution is $\gamma^{\dagger}_{0,-}=(0,1,-1,0)^{T}\exp\left(-\int^r_0 dr'(\Delta_0(r')-g_s)/v_F\right)$. This represents MZM is from the consistent phase winding between $\Delta(\mathbf{r})$ and magnetic texture $\boldsymbol{\mathcal{A}}_1$.

Now we understand that the radial magnetic texture effectively act like an Abrikosov vortex, where the Abrikosov vortex core is at the center of magnetic texture. We put the Dirac surface state on a sphere, and assume magnetic vortex ($g_s>0$) and anti-vortex ($g_s<0$) pairs are located at the north and south poles, respectively. In Fig.~\ref{fig3}(a), when the flux line of the effective Abrikosov vortex penetrates the two poles, the pairing potential is $\Delta_0(r)e^{i\theta}$ and $\Delta_0(r)e^{-i\theta}$ locally at the north and south poles, respectively. The solutions for MZM at north and south poles are exactly the same, for they are the time-reversal partner of each other. Now in the case of weak exchange interaction $g_s$, where the radial magnetic texture could not contribute to a flux quantum. We need add external magnetic field to generate Abrikosov vortex. If the external generated Abrikosov vortex core coincides with the center of the magnetic texture, the MZM locates at the north and south poles. Now the flux line is adiabatically shifted away from the south pole to point $Z$ in Fig.~\ref{fig3}(b). For the condition when the BdG spectrum has a full gap with a vortex-free pairing potential (which is $g_s<\Delta_0$ illustrated in SM~\cite{supple}), then the MZM, if exist, can only be localized at the Abrikosov vortex core, since away from the vortex core the surface spectrum is gapped. And the condition for MZM existence in the vortex core at $Z$ point should be the same as that at north pole ($g_s<\Delta_0$), otherwise it will be contradictory to the fact that MZMs always come into pairs. This arguement can also be understood in the limit when $Z$ point is far away from the origin of magnetic textures, then the magnetization is approximately uniform and parallel to the surface. The parallel in-plane exchange term shift the Dirac point away from $\Gamma$ point in the perpendicular direction, and introduce pairing-breaking effect between states at $\mathbf{k}$ and $-\mathbf{k}$. The superconducting surface states remains topological with a gap when $\Delta_0>g_s$~\cite{yuan2018}, which is exactly the condition for the existence of MZMs in the vortex core. However, the adiabatical continuity fails when the BdG spectrum is gapless with a vortex-free pairing potential. 

\begin{figure}[t]
\begin{center}
\includegraphics[width=3.4in,clip=true]{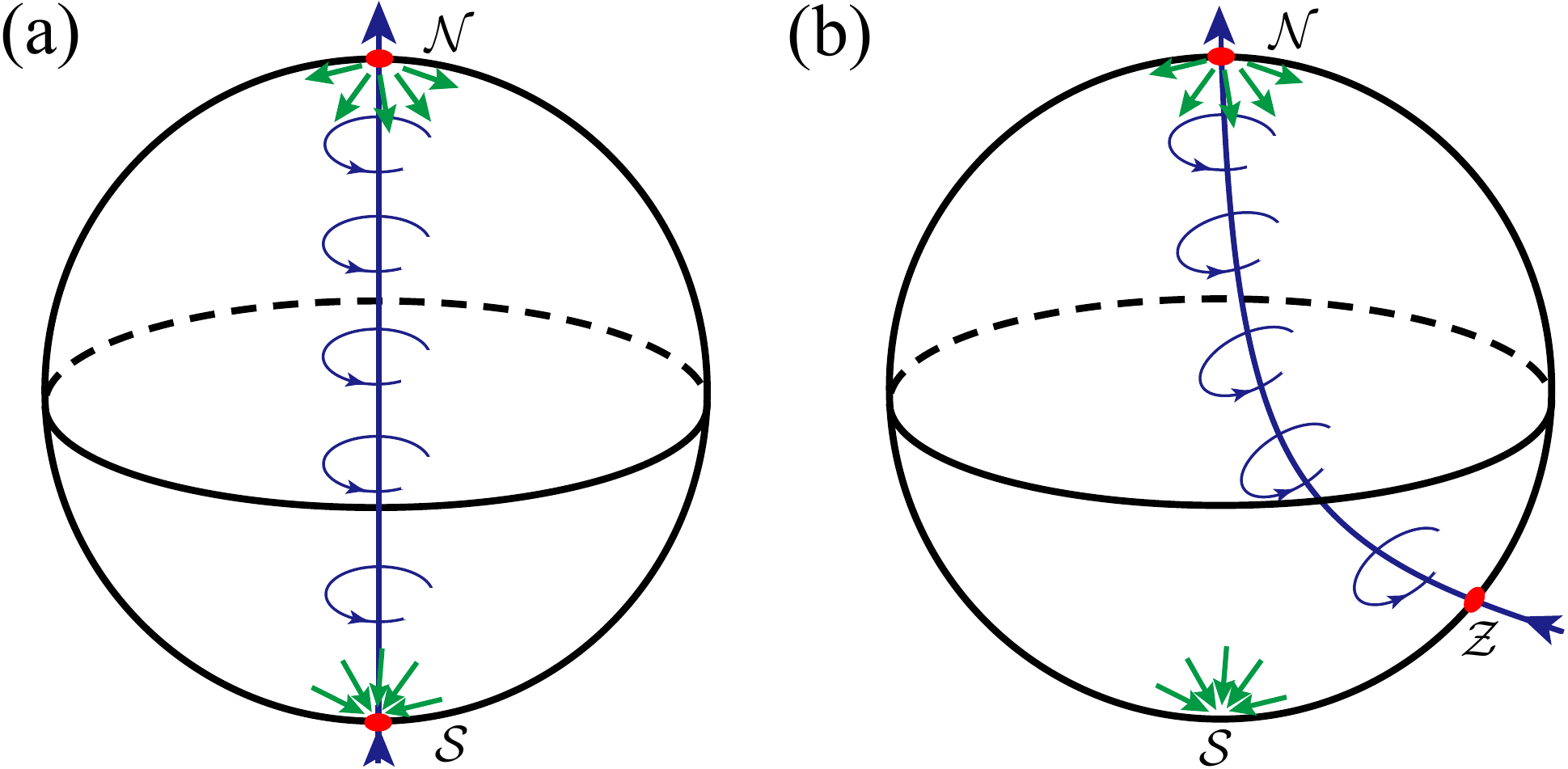}
\end{center}
\caption{(a,b) The MZMs are trapped near the surface in the Abrikosov vortex core. The vortex center coincides with the center of magnetic texture in (a) but is shifted away in (b).}
\label{fig3}
\end{figure}

\emph{Physical realization.-} 
The above 2D magnetic texture has a singularity at the origin, while in ferromagnet films with a hard axis pointing along $z$-axis, a magnetic vortex with spiralling magnetization configuration could emerge. $\mathbf{n}(\mathbf{r})=(\cos\varphi\sqrt{1-n_z^2},\sin\varphi\sqrt{1-n_z^2},n_z)$, where $n_z(r)$ depends only on $r=\sqrt{x^2+y^2}$, $\varphi=q\phi+\phi_0$. The radial vortex $q=1, \phi_0=0$ is stablized interfacial Dzyaloshinskii-Moriya interaction (DMI)~\cite{finocchio2016}, while curling vortex $q=0, \phi_0=-\pi/2$ is obtained by minimizing the dipolar interactions~\cite{shinjo2000,wachowiak2002}. They are the same as that in Fig.~\ref{fig1}(a,b) at large $r$, respectively. 
Generally $n_z(0)=1$ and decays to $n(\xi_0)=0$ within the decay length $\xi_0$, a typical form is $n_z(r)=\text{sech}(r/\xi_0)$ with $\xi_0\approx5\sim10$~nm. 
As long as $a>\xi_0$, the above study applies to the coupling between Dirac electron and such a realistic magnetic vortex. Here $a$ is the size of the magnetic vortex within which a flux quantum is obtained. The bound state located at $r_{\text{max}}$ may be probed by the scanning tunnelling microscopy. For an estimation, take $v_F\approx5\times10^5$~m/s in Bi$_2$Te$_3$, then $a\approx30$~nm when $g_s=11$~meV.

We propose the artificial two bilayer ferromagnets heterostructure to create magnetic radial vortex, where the dipolar energy is minimized by an antiferromagnetic coupling between them and the radial vortex is obtained by the DMI gradient along the radial direction, namely the DMI parameter monotonically decreases as $r$ increases.

The incorporation of the magnetic proximity effect into a TI has been exemplified in heterostructures with magnetic insulators~\cite{wei2013,katmis2016,tang2017}. Recently a van der Waals magnetic TI MnBi$_2$Te$_4$ and its descendents have been discovered~\cite{zhang2019,otrokov2019,gong2019,wuj2019,yan2020,hu2020}, which is compatible with the Bi$_2$Te$_3$ family materials. Such experimental progress on the material growth and rich material choice of TI and magnetic insulators makes it possible to realize the magnetic vortex in TI heterstructures.

\begin{acknowledgments}
We acknowledge Jiang Xiao for valuable discussions. This work is supported by the National Key Research Program of China under Grant No.~2019YFA0308404, the Natural Science Foundation of China through Grant Nos.~11774065 and~12174066, Science and Technology Commission of Shanghai Municipality under Grant No.~20JC1415900, and the Natural Science Foundation of Shanghai under Grant No.~19ZR1471400. C. W. is supported by the Natural Science Foundation of China through Grant No.~12174317, and No.~11729402.

\end{acknowledgments}

\begin{widetext}
\section*{Supplementary Material for "Planar magnetic texture on the surface of a topological insulator"}

\section{S1. NUMERICAL METHOD TO DIRAC FERMION COUPLED TO MAGNETIC TEXTURES}
In this section, we provide the detail of numerical calculation method for Dirac fermion with 
Hamiltonian given by:
\begin{equation}
    \begin{aligned}
      H_0&=v_F((\mathbf{k}-\boldsymbol{\mathcal{A}})\times\boldsymbol{\sigma})\cdot\hat{\mathbf{e}}_z\\
        &=v_F(k_x-A_x)\sigma_y-v_f(k_y-A_y)\sigma_x\\
        &=\left(                
        \begin{array}{cc}   
          0 & -v_F(\partial_{x}-i\partial_{y})+v_FA(r,\phi)\\   
          v_F(\partial_{x}+i\partial_{y})+v_FA^{*}(r)  & 0\\   
        \end{array}
        \right) 
    \end{aligned}
\end{equation}
with $A(r,\phi)=(g_s/{v_F})(-iy/r+x/r)=(g_s/{v_F})e^{-i\phi}$ for radial vortex and $A(r,\phi)=(g_s/v_F)(ix/r+y/r)=
i(g_s/v_F)e^{-i\phi}$ for curling vortex. We numerically solve this equation in a disk geometry.
Cause $\partial_x\pm i\partial_y=e^{i\pm\phi}(\partial_r\pm i\partial_{\phi}/r)$, the wave function can be factorized as 
\begin{equation}
  \psi_{n,l}=e^{i l\phi}(u_{n,l,\uparrow},u_{n,l+1,\downarrow}e^{i\phi})^T
\end{equation}
and the radial wave function can be expanded in terms of Bessel function of the first kind
\begin{equation}
  u_{n,l}=\sum_{j=1}^N c_{n,j,l}\frac{\sqrt{2}}{RJ_{l+1}(\beta_{j,l})}J_{l}(\beta_{j,l}\frac{r}{R})\equiv\sum_{j=1}^N 
  c_{n,j,l}\phi_{j,l}
\end{equation}
where $J_l$ is m-th order Bessel function of the first kind, $\beta_{j,l}$ is its j-th zeros, N is cutoff for the 
expansion and $R$ is the radius of our disk. Then the Hamiltonian is reduced to $2N\times 2N$ matrix as 
\begin{equation}
  H_{BdG}=\left(                
      \begin{array}{cc}  
        0 & V_{\mu,\mu+1}+A_{\mu,\mu+1}\\ 
        V^{T}_{\mu,\mu+1}+A^{\dagger}_{\mu,\mu+1}  & 0 \\ 
      \end{array}
      \right)
\end{equation}
with 
\begin{equation}
  (V_{l,l'})_{i,j}=\frac{2v_F}{R}\frac{\beta_{i,l}\beta_{j,l'}}{\beta^2_{i,l}-\beta^2_{j,l'}}\quad 
  (A_{l,l'})_{i,j}=\int^{R}_{0}rdr A(r)\phi_{i,l}\phi_{j,l'}
\end{equation}
In the following calculations we choose $v_F=10,R=400,N=400$ which is enough to ensure the convergence of the low 
energy spectrum.
\section{S2. NUMERICAL METHOD TO DIRAC FERMION BdG HAMILTONIAN}
In presence of pairing vortex $\Delta(r)=\Delta_0\tanh(r/\epsilon_0)e^{i\phi}$, the wave function is given by 
\begin{equation}
  \Phi_{n,l}=e^{il\phi}(u_{n,l,\uparrow},u_{n,l+1,\downarrow}e^{i\phi},v_{n,l-1,\downarrow}e^{-i\phi},-v_{n,l,\uparrow})^T
\end{equation}
Following the same method, we expand the radial part by Bessel functions,  
then the BdG matrix is 
\begin{equation}
  H_{BdG}=\left(                 
  \begin{array}{cccc}   
    \Delta_z-\mu & V_{l,l+1}+A_{l,l+1} &\Delta_{l,l-1}& 0  \\ 
    V^{T}_{l,l+1}+A^{\dagger}_{l,l+1}  & -\Delta_z-\mu & 0& \Delta_{l+1,l} \\ 
    \Delta^{T}_{l,l-1} & 0 & \Delta_z+\mu & -V_{l-1,l}+A_{l-1,l}  \\  
      0 & \Delta^{T}_{l+1,l}  &  -V^{T}_{l-1,l}+A^{\dagger}_{l-1,l}        &-\Delta_z+\mu   \\  
  \end{array}
  \right)  
\end{equation}
with $\mu$ as the Chemical potential and $\Delta_z$ as the z direction Zeeman field.
 
As we can see from Fig.~\ref{Fig_BdG_Vortex}, the zero mode doesn't split with a nonzero Zeeman field when $g_s<\Delta_0$. But 
zero modes in Fig.~\ref{Fig_BdG_Vortex}(b) deviate from zero which indicates they are originated from bound states in Fig.~1(c) of
the main text.
\begin{figure}[H]
  \begin{center}
  \includegraphics[width=4.0in,clip=true]{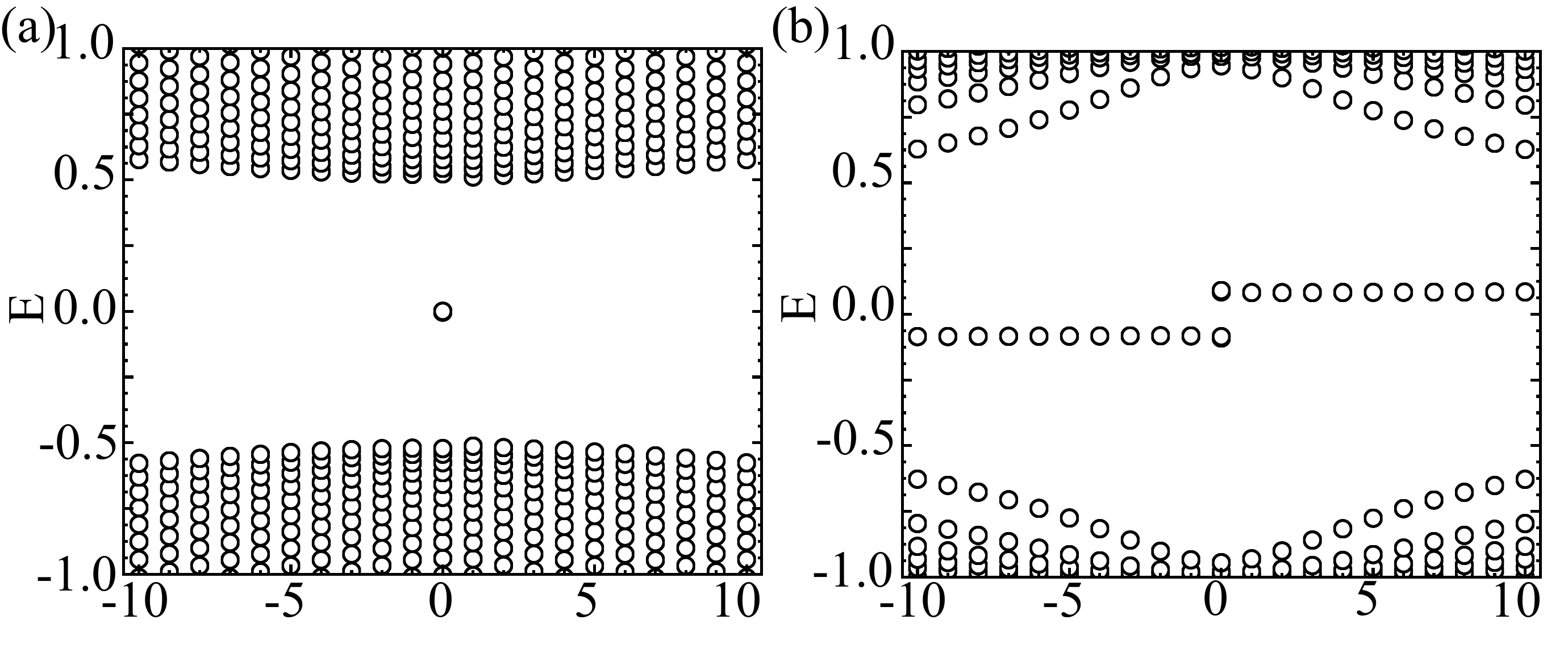}
  \end{center}
  \caption{The BdG spectrum of the Abrikosov vortex $\Delta_0(r)e^{i\phi}$ in presence of the z-direction 
  Zeeman field for the radial vortex and $g_s$=0.5/2 for (a,b).
  Other parameters are $v_F=10,\Delta_0=1$ and $\Delta_z=0.1$.}
  \label{Fig_BdG_Vortex}
\end{figure}
Next, lets consider the the case with $g_s<0$. The phenomena is similar with $g_s>0$ as the majorana zero modes exists only when 
$|g_s|>|\Delta_0|$. And nonzero chemical potential would make the system gapless when $|g_s|>|\Delta_0|$.
\begin{figure}[H]
  \begin{center}
  \includegraphics[width=7.0in,clip=true]{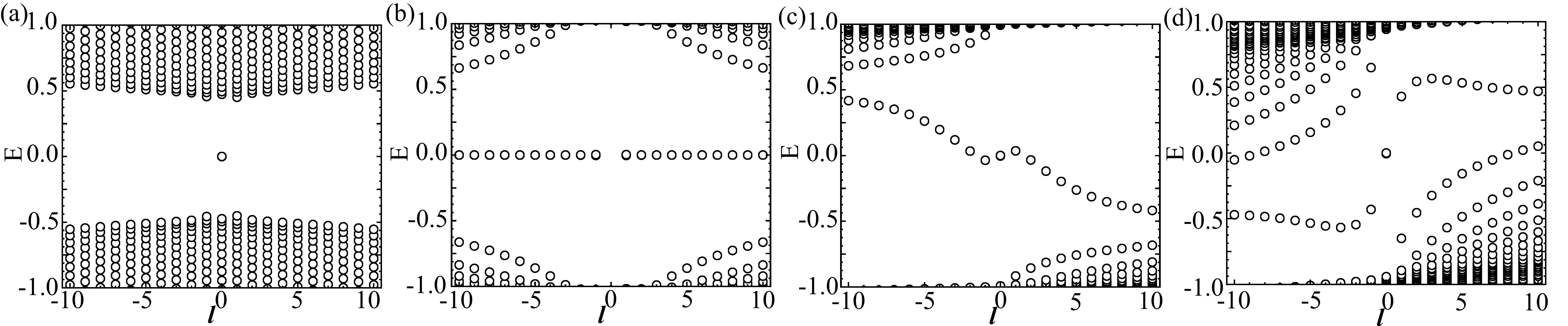}
  \end{center}
  \caption{The BdG spectrum of the Abrikosov vortex $\Delta_0(r)e^{i\phi}$ with $g_s<0$ for
  the radial vortex. (a) $g_s=-0.5, \mu=0$. (b) $g_s=-2, \mu=0$. (c) $g_s=-0.5, \mu=5$.
  (d) $g_s=-1.5, \mu=5$.
  Other parameters are $v_F=10,\Delta_0=1$.}
  \label{Fig_BdG_Vortex_2}
\end{figure}

\section{S3. Vortex free BdG spectrum}
 
Finally, we present the BdG spectrum for vortex free superconductor pairing. In this case, the superconductor pairing is 
$\Delta(r)=\Delta_0\tanh(r/\epsilon_0)$. Due to absence of winding number, the wave should factorized as 
\begin{equation}
  \Phi_{n,l}=e^{il\phi}(u_{n,l,\uparrow},u_{n,l+1,\downarrow}e^{i\phi},v_{n,l,\downarrow}e^{-i\phi},-v_{n,l+1,\uparrow})^T
\end{equation}
such that the angular momentum is $j=l+1/2$ which implies the absence of the majorana zero mode. 
Fig.~\ref{Fig_BdG_Vortex_3} shows the BdG spectrums for radial. Similar to the Abrikosov
votrex case, the system is always gapped when $g_s<\Delta_0$.
\begin{figure}[H]
  \begin{center}
  \includegraphics[width=7.0in,clip=true]{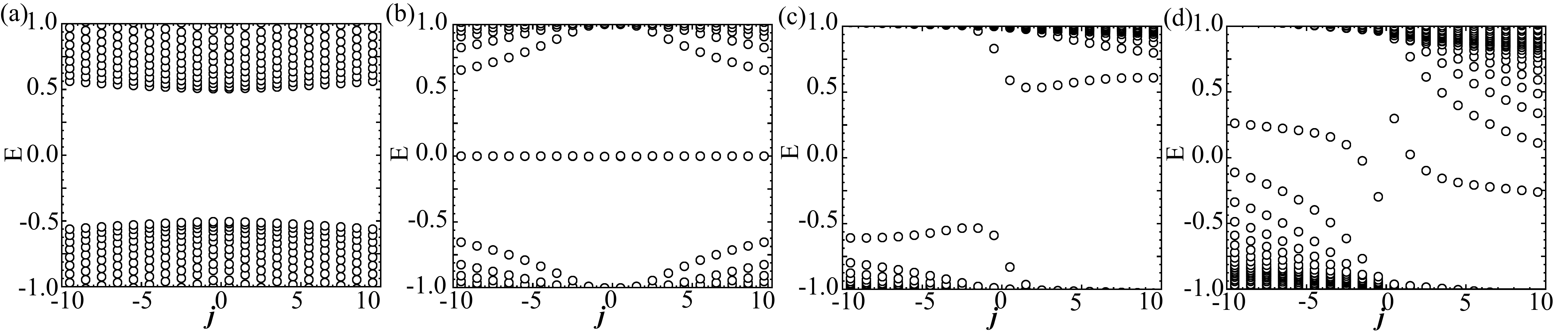}
  \end{center}
  \caption{The BdG spectrum of the Abrikosov vortex $\Delta_0(r)$ for
  the radial vortex. (a) $g_s=0.5, \mu=0$. (b) $g_s=2, \mu=0$. (c) $g_s=0.5, \mu=5$.
  (d) $g_s=1.5, \mu=5$.
  Other parameters are $v_F=10,\Delta_0=1$.}
  \label{Fig_BdG_Vortex_3}
\end{figure}
\end{widetext}

\end{document}